\documentclass[slac_one]{revtex4}
\usepackage{graphicx}
\usepackage{fancyhdr}
\newcommand{\be}{\begin{equation}}
\newcommand{\ee}{\end{equation}}
\newcommand{\bea}{\begin{eqnarray}}
\newcommand{\eea}{\end{eqnarray}}
\def\gapp{\mathrel{\raise.3ex\hbox{$>$}\mkern-14mu
              \lower0.6ex\hbox{$\sim$}}}
\def\gsim{\gapp}
\def\lapp{\mathrel{\raise.3ex\hbox{$<$}\mkern-14mu
              \lower0.6ex\hbox{$\sim$}}}
\def\lsim{\lapp}
\newcommand{\vp}{\varphi}
\newcommand{\mpl}{m_p}

\def\mstop{m_{\,\widetilde{t}}}

\begin{document}

\title{Baryogenesis and Leptogenesis}

\author{Mark Trodden}
\affiliation{Department of Physics, Syracuse University, 
Syracuse, NY 13244-1130, USA.}

\begin{abstract}
The energy budget of the universe contains two components, dark matter and dark energy, about which
we have much to learn. One should not forget, however, that the baryonic component presents its own
questions for particle cosmology. In the context of cosmology, baryons would have annihilated with their antiparticles in the early universe, leaving a negligible abundance of baryons, in disagreement with that
observed. In this general lecture, delivered at the SLAC 2004 Summer Science Institute, I provide an
overview of the central issue and the general principles behind candidate models. I also briefly discuss
some popular examples of models that are firmly rooted in particle physics.
\end{abstract}

\maketitle

\thispagestyle{fancy}

\section{Introduction}
\label{Intro}
The audience for this talk was extremely diverse, ranging from beginning graduate students, through experts in subfields of physics somewhat distinct from the subject matter of my talk, to baryogenesis experts. My brief was to present an overview of the main issues facing baryogenesis, accessible to all members of the audience. The original talk was an hour long, and I attempted to stay true to the organizers' requests regarding the level. In writing up these proceedings, I have taken the opportunity to flesh out some of the topics with considerably more detail, drawing heavily from my two review articles~\cite{Riotto:1999yt,Trodden:1998ym}.

The problem of the baryon asymmetry of the universe is a classic problem of particle cosmology. Particle physics has taught us that matter and antimatter behave essentially identically, and indeed the interactions between matter and antimatter are the focus of successful terrestrial experiments. On the other hand, cosmology teaches us that the early universe was an extremely hot, and hence energetic, environment in which one would expect equal numbers of baryons and antibaryons to be copiously produced. This early state of the universe stands in stark contrast to what we observe in the universe today. Direct observation shows that the universe around us contains no appreciable primordial antimatter. In addition, the theory of primordial nucleosynthesis (for a review see \cite{CST 95}) allows accurate predictions of the cosmological
abundances of all the light elements, H, $^3$He, $^4$He, D, B
and $^7$Li, while requiring only that, defining $n_{b({\bar b})}$ to be the number density of (anti)-baryons and $s$ to be the entropy density,
\begin{equation}
2.6\times 10^{-10} < \eta\equiv \frac{n_b -n_{\bar b}}{s} < 6.2\times 10^{-10} \ ,
\end{equation}
(see, for example,~\cite{Fields:cn}).
Very recently this number has been independently determined to be $\eta =  6.1\times 10^{-10}\ ^{+0.3\times 10^{-10}}_{-0.2\times 10^{-10}}$ from precise measurements of the relative heights of the first two microwave background (CMB) acoustic peaks by the WMAP satellite~\cite{Bennett:2003bz}.
Alternatively we may 
write the range as
\be
0.015(0.011)\lsim \Omega_B\:h^2\lsim 0.026 (0.038) \ ,
\ee
where $\Omega_B$ is the proportion of the critical energy density in
baryons, and $h$ parametrizes the present value of the
Hubble parameter via $h=H_0/(100$ Km Mpc$^{-1}$ sec$^{-1})$.

The standard cosmological model cannot explain the observed
value of $\eta$. To see this, suppose that initially we start with $\eta=0$. 
At temperatures $T\lsim$ 1 GeV the equilibrium abundance of 
nucleons and antinucleons is
\be
\label{a}
\frac{n_b}{n_\gamma}\simeq\frac{n_{{\bar b}}}{n_\gamma}\simeq
\left(\frac{m_p}{T}\right)^{3/2}\:e^{-\frac{m_p}{T}}.
\ee
As the universe cools, the number of nucleons and antinucleons 
decreases as long as the annihilation rate 
$\Gamma_{{\rm ann}}\simeq n_b\langle \sigma_A v\rangle$ is larger than the 
expansion rate of the universe $
H\simeq 1.66\:g_*^{1/2}\frac{T^2}{\mpl}$. The thermally averaged annihilation 
cross section $\langle \sigma_A v\rangle$ is of the order of $m_\pi^2$, so 
at $T\simeq$ 20 MeV, $\Gamma_{{\rm ann}}\simeq H$, and annihilations freeze 
out, nucleons and antinucleons being so rare that they cannot annihilate any 
longer. Therefore, from (\ref{a}) we obtain
\be
\frac{n_b}{n_\gamma}\simeq \frac{n_{{\bar b}}}{n_\gamma}\simeq 10^{-18},
\ee
which is much smaller than the value required by nucleosynthesis. 

An initial 
asymmetry may be imposed by hand as an initial condition, but this would 
violate any naturalness principle. Rather, the goal of this talk was to describe recent progress in understanding scenarios for 
generating the baryon asymmetry of the universe (BAU) within the context of modern cosmology.
Much more complete reviews of this subject, with several different approaches, can be found 
in~\cite{Riotto:1999yt,Trodden:1998ym,Dine:2003ax,review,NTreview,CKNreview,AD 92}.  

In the next section I describe the {\it Sakharov Criteria} which must
be satisfied by any particle physics theory through which a baryon
asymmetry is produced. In section~(\ref{GUT}) I briefly describe how baryogenesis can
proceed in certain Grand Unified Theories and mention how new ideas concerning 
preheating after inflation have modified this
scenario. I also briefly discuss the idea of leptogenesis. In section~(\ref{EWBG}) I
review at length one of the most popular models - electroweak baryogenesis - and in section~(\ref{AD})
I briefly mention Affleck-Dine (AD) type baryogenesis scenarios. The choices of what topics to cover and 
how much detail to devote to each were decided by a combination of personal taste and 
the extent of the direct connection to current and upcoming collider experiments.

Throughout I use a metric with signature $+2$
and, unless explicitly stated otherwise, I employ units such that
${\hbar}=c=k=1$.

\section{The Sakharov Criteria}
\label{sakharov}

As pointed out by Sakharov \cite{sak}, a small baryon asymmetry $\eta$ 
may have been produced in the early universe if three  necessary conditions 
are satisfied: {\it i)} baryon number ($B$) violation; {\it ii)} violation of 
$C$ (charge conjugation symmetry) and $CP$ (the composition of parity and 
$C$) and {\it iii)} departure from thermal equilibrium. The first 
condition should be clear since, starting from a  baryon symmetric universe 
with $\eta=0$, baryon number violation must take place in order to  evolve 
into  a universe in which $\eta$ does not vanish. The second Sakharov 
criterion is required 
because, if $C$ and $CP$ are exact symmetries, then one can prove
that the total rate for any process which produces an excess of baryons is
equal to the rate of the complementary process which produces an
excess of antibaryons and so no net baryon number can be created. 
That is to say that the thermal average of the baryon number operator $B$, 
which is odd under
both $C$ and $CP$, is zero unless those discrete symmetries are
violated. $CP$-violation is present either if  there are complex phases in 
the lagrangian which cannot be reabsorbed by field redifinitions (explicit 
breaking) or if some Higgs scalar field acquires a VEV which is not real 
(spontaneous breaking).

Finally, to explain the third criterion, one can calculate the equilibrium 
average of $B$
\bea
\langle B\rangle_T & = & {\rm Tr}\:(e^{-\beta H}B)=
 {\rm Tr}\:[(CPT)(CPT)^{-1}e^{-\beta H}B)] \nonumber \\
& = & {\rm Tr}\:(e^{-\beta H}(CPT)^{-1}B(CPT)]=
-{\rm Tr}\:(e^{-\beta H}B) \ ,
\eea
where I have used that the Hamiltonian $H$ commutes with $CPT$. Thus
$\langle B\rangle_T = 0$ in equilibrium and there is no generation of
net baryon number.

Of the three Sakharov conditions, baryon number violation and $C$ and 
$CP$-violation may be investigated  only within a given particle physics 
model, while the third condition -- the departure from thermal equilibrium -- 
may be discussed in a more general way, as we shall see.  
Let us discuss the Sakharov criteria in more detail.

\subsection{Baryon Number Violation}

\subsubsection{$B$-violation in Grand Unified Theories}
Baryon number violation is very natural in Grand Unified Theories (GUTs)~\cite{lan}, since
a general property is that the same representation of the gauge group
$G$ may contain both 
quarks and leptons, and therefore it is possible for scalar and gauge bosons 
to mediate gauge interactions among fermions having different baryon 
number. 
However, this alone is not sufficient to conclude that baryon number is 
automatically  violated in GUTs, since in some circumstances it is 
possible to assign a baryonic charge to the gauge bosons in such a way 
that at each boson-fermion-fermion vertex the baryon number is conserved. 
For example, in the particular case of the gauge group $SU(5)$, it turns out that among 
all the scalar and gauge bosons which couple only to the fermions 
of the SM, only five of them may give rise to interactions which violate the 
baryon number. 

\subsection{$B$-violation in the Electroweak Theory.}

It is well-known that the most general Lagrangian invariant 
under the SM gauge group and containing only color singlet Higgs fields 
is automatically invariant under global abelian 
symmetries which may be identified with the baryonic and leptonic symmetries. 
These, therefore, are accidental symmetries and as a result it is not 
possible to violate $B$ and $L$ at tree-level or at any order of 
perturbation theory. Nevertheless, in many cases the perturbative expansion 
does not describe all the dynamics of the theory and, indeed, in 1976 't 
Hooft \cite{ho}
realized that nonperturbative effects (instantons) may give rise to processes 
which violate the combination $B+L$, but not the orthogonal combination 
$B-L$. The probability of these processes occurring today is exponentially 
suppressed and probably irrelevant. However, in more extreme situations -- 
like the primordial universe at very high temperatures 
\cite{ds,manton,km,krs} -- baryon and lepton number violating processes 
may be fast enough to play a significant role in baryogenesis. Let us have 
a closer look.

At the quantum level, the baryon and the lepton symmetries are  
anomalous \cite{adler,bj}
\be
\partial_{\mu}j_B^{\mu} = \partial_{\mu}j_L^{\mu}
=n_f\left(\frac{g^2}{32\pi^2}W_{\mu\nu}^a {\tilde W}^{a\mu\nu}
-\frac{g'^2}{32\pi^2}F_{\mu\nu}{\tilde F}^{\mu\nu}\right) \ ,
\label{anomaly}
\ee
where $g$ and $g'$ are the gauge couplings of $SU(2)_L$ and $U(1)_Y$, 
respectively,  $n_f$ is the number of families and $
\tilde{W}^{\mu\nu} = (1/2) \epsilon^{\mu\nu\alpha\beta}W_{\alpha\beta}$ 
is the dual of the $SU(2)_L$ field strength tensor, with an analogous
expression holding for ${\tilde F}$. To understand how the anomaly is closely 
related to the vacuum structure of the theory, we may compute  the change in
baryon number from time $t=0$ to some arbitrary final time $t=t_f$.
For transitions between vacua, the average values of the field strengths 
are zero at the beginning and the end of the evolution. The change in baryon 
number may be written as

\be
\Delta B = \Delta N_{CS} \equiv n_f[N_{CS}(t_f) - N_{CS}(0)]\ .
\ee
where the Chern-Simons number is defined to be  

\begin{equation}
N_{CS}(t) \equiv \frac{g^2}{32 \pi^2}\int d^3x\, \epsilon^{ijk}
\:{\rm Tr}\:\left( A_i \partial_j A_k + \frac{2}{3}ig A_i A_j A_k \right)\ .
\label{NCSdef}
\end{equation}
Although the Chern-Simons number is not gauge invariant, the change 
$\Delta N_{CS}$ is. Thus, changes in Chern-Simons number result in 
changes in baryon number which are integral multiples of the number of
families $n_f$. Gauge transformations $U(x)$ which connects two degenerate 
vacua of the gauge theory  may  change the Chern-Simons number by an integer 
$n$, the winding number. If the system is able to perform a transition from 
the vacuum ${\cal G}_{{\rm vac}}^{(n)}$ to the closest one
${\cal G}_{{\rm vac}}^{(n\pm 1)}$, the Chern-Simons number is changed by 
unity and $\Delta B=\Delta L=n_f$. 
Each transition creates 9 left-handed quarks (3 color states for each 
generation) and 3 left-handed leptons (one per generation). 
 However, adjacent vacua of the electroweak theory are separated by a ridge of
configurations with energies larger than that of the vacuum.  
The lowest energy point on this ridge is a saddle point solution
to the equations of motion with a single negative eigenvalue, and
is referred to as the {\it sphaleron} \cite{manton,km}.
The probability of baryon number nonconserving processes at zero temperature
 has been computed by 't Hooft \cite{ho} and is highly suppressed by a 
factor  ${\rm exp}(-4\pi/ \alpha_W)$, where $\alpha_W=g^2/4\pi$. This factor 
may be interpreted as the probability of making a transition from one 
classical vacuum to the closest one by tunneling through an energy 
barrier of height $\sim 10$ TeV corresponding to the sphaleron. On the 
other hand,  
one might think that fast  baryon  number violating transitions may be 
obtained in physical situations which involve a large number of fields. 
Since the sphaleron may be produced by collective and coherent excitations 
containing $\sim  1/\alpha_W$ quanta with wavelength of the order of $1/M_W$, 
one expects that  at temperatures $T\gg M_W$, these modes essentially obey 
statistical mechanics and the transition probability may be computed via 
classical considerations. 
 The general 
framework for evaluating the thermal rate of anomalous processes in the
electroweak theory was developed in \cite{krs}.
 Analogously to the case of zero temperature
and since the transition which violates the baryon number
is   sustained by the sphaleron configuration, the thermal rate
of baryon number violation in the {\it broken} phase  is proportional to 
${\rm exp}(-S_3/T)$, where $S_3$ is the three-diemensional action computed 
along the sphaleron configuration, 

\be
S_3 = E_{sp}(T) \equiv (M_W(T)/\alpha_W) {\cal E} \ ,
\ee 
with the dimensionless parameter ${\cal E}$ lying in the range $
3.1 < {\cal E} < 5.4$ 
depending on the Higgs mass.
The prefactor of the thermal rate  was computed in \cite{carson} as
\be
\Gamma_{sp}(T) = \mu\left(\frac{M_W}{\alpha_W T}\right)^3M_W^4 
\exp\left(-\frac{E_{sph}(T)}{T}\right) \ ,
\label{brokenrate}
\ee
where $\mu$ is a dimensionless constant.
Recent approaches to
calculating the rate of baryon number violating events in the broken
phase have been primarily numerical. 
The best 
calculation to date of the broken phase sphaleron rate is undoubtably
that by Moore \cite{moore}. This work yields a fully nonperturbative 
evaluation of the broken phase rate by using a combination of multicanonical
weighting and real time techniques.

Although the Boltzmann suppression in~(\ref{brokenrate}) appears large,
it is to 
be expected that, when the electroweak symmetry becomes restored  at a 
temperature of around $100\,$GeV, there will no longer be an exponential 
suppression factor. Although calculation of the baryon number violating 
rate in the 
high temperature {\it unbroken} phase is extremely difficult, a simple 
estimate is 
possible. The only important scale in the symmetric phase is the 
magnetic screening length given by $
\xi=(\alpha_W T)^{-1}$.
Thus, on 
dimensional grounds, we 
expect the rate per unit volume of sphaleron events to be

\be
\Gamma_{sp}(T)=\kappa(\alpha_W T)^4 \ ,
\label{unbrokenrate}
\ee
with $\kappa$ another dimensionless constant. The rate of sphaleron processes
can be related to the diffusion constant for Chern-Simons 
number  by a fluctuation-dissipation theorem
\cite{ks88} (for a good description of this see \cite{review}). In
almost all numerical calculations of the sphaleron rate, this relationship 
is used and what is actually evaluated is the diffusion constant.

The simple scaling argument leading to (\ref{unbrokenrate}) does not capture all of the
important dynamics~\cite{AAPS90,AAPS91,ambjorni,GM96,ambjorniii,M&Tii 97,arnold1,arnold2}. The effective dynamics of soft nonabelian 
gauge fields at finite temperature has been addressed in 
\cite{bod,arlog}, where it was found that 
$\Gamma_{sp}\sim \alpha_W^5 T^4\:{\rm ln}(1/\alpha_W)$. Further,
Lattice simulations with hard-thermal loops included~\cite{mooremu} indicate
 $\Gamma_{sp}\sim 30\alpha_W^5 T^4$.

\subsubsection{Other Ways of Achieving $B$-Violation.}

One of the features which distinguishes supersymmetric
field theories \cite{nilles,haber} from ordinary ones is the existence of
``flat directions" in field space on which the scalar potential vanishes.
At the level of renormalizable terms, such flat directions
are generic.
Supersymmetry breaking lifts these directions and
sets the scale for their potential.
>From a cosmological perspective, these flat directions
can have profound consequences.  The parameters which
describe the flat directions can be thought of as expectation
values of massless chiral fields, the moduli.
Many flat directions  present
in the minimal supersymmetric standard model (MSSM), 
 carry baryon or lepton
number, since  along them  squarks
and sleptons have non zero vacuum expectation values (VEVs). 
If baryon and lepton number are explicitly broken
it is possible to excite a non-zero baryon or lepton number along such
directions, as first suggested by Affleck and Dine \cite{affleckdine} in a more
general context. I'll explore these ideas more thoroughly in section 
\ref{AD}.

\subsection{$CP$-Violation}

\subsubsection{$CP$-Violation in Grand Unified Theories.}

$CP$-violation in GUTs arises in loop-diagram corrections to 
baryon number violating bosonic decays. Since it is necessary that 
the particles in the loop also undergo 
$B$-violating decays, the relevant particles are the $X$, $Y$, and 
$H_3$ bosons in the case of $SU(5)$. In that case, 
$CP$-violation is due to the complex phases of the Yukawa couplings $h_U$ and 
$h_D$ which cannot be reabsorbed by field redefinition. At tree-level 
these phases do not give any contribution to the baryon asymmetry and at the 
one-loop level the asymmetry is proportional to $
{\rm Im}\:{\rm Tr}\:\left(h_U^\dagger h_U h_D^\dagger h_D\right)=0$,
where the trace is over generation indices. 
The problem of too little $CP$-violation in $SU(5)$ may be solved by further
complicating the Higgs sector.  For example, adding a Higgs in the 
$45$ representation of $SU(5)$ leads to an adequate baryon asymmetry 
for a wide range of the parameters \cite{harvey1}. In the case of $SO(10)$, 
$CP$-violation may be  provided by the complex Yukawa couplings between the 
right-handed and the  left-handed neutrinos and the scalar Higgs. 

\subsubsection{$CP$-Violation in the CKM Matrix of the Standard Model.}

Since  only the left-handed 
electron is $SU(2)_L$ gauge coupled, $C$ is maximally broken in the SM.  
Moreover, $CP$ is known not to be an exact symmetry
of the weak interactions. This is seen experimentally in the neutral 
Kaon system through $K_0$, ${\bar K}_0$ mixing. At
present there is no accepted theoretical explanation of this.
However, it is true that $CP$-violation is a natural feature of the
standard electroweak model. There exists a
charged current which, in the weak interaction basis, may be written
as $
{\cal L}_W = (g/\sqrt{2}) {\bar U}_L \gamma^{\mu} D_L W_{\mu}
+ {\rm h.c.}$, 
where $U_L=(u,c,t,\ldots)_L$ and $D_L=(d,s,b,\ldots)_L$. Now, the
quark mass matrices may be diagonalized by unitary matrices
$V^U_L$, $V^U_R$, $V^D_L$, $V^D_R$ via
\bea
{\rm diag}(m_u,m_c,m_t,\ldots) & = & V^U_L M^U V^U_R \ ,  \\
{\rm diag}(m_d,m_s,m_b,\ldots) & = & V^D_L M^D V^D_R \ .
\eea
Thus, in the basis of quark mass eigenstates, the charged current may be 
rewritten as $
{\cal L}_W = (g/\sqrt{2}){\bar U}_L'K\gamma^{\mu}D_L' W_{\mu}
+{\rm h.c.}$, 
where $U_L'\equiv V^U_L U_L$ and $D_L'\equiv V^D_L D_L$. The matrix
$K$, defined by $
K \equiv V_L^U (V_L^D)^{\dagger}$, 
is referred to as the Kobayashi-Maskawa (KM) quark mass mixing
matrix. 
For three generations, as in the
SM, there is precisely one independent nonzero phase $\delta$ which signals
$CP$-violation.
While this is
encouraging for baryogenesis, it turns out that this particular source of
$CP$-violation is not strong enough. The relevant effects are parameterized by
a dimensionless constant which is no larger than $10^{-20}$. This appears
to be much too small to account for the observed BAU and, thus far, 
attempts to utilize this source of $CP$-violation for electroweak
baryogenesis have been unsuccessful. In light 
of this, it is usual to extend the SM in some minimal 
fashion that increases the amount of $CP$-violation in the theory while not
leading to results that conflict with current experimental data.

\subsubsection{$CP$-Violation in Supersymmetric Models}

The most promising and  
well-motivated framework incorporating $CP$-violation beyond the SM 
seems to be supersymmetry  \cite{nilles,haber}. 
Let us consider the MSSM superpotential
\be
\label{superpot}
W=\mu\hat{H}_1\hat{H}_2+h^u\hat{H}_2\hat{Q}\hat{u}^c+h^d\hat{H}_1
\hat{Q}\hat{d}^c+h^e\hat{H}_1\hat{L}\hat{e}^c,
\ee
where we have omitted the generation indices. Here $\hat{H}_1$ and $\hat{H}_2$
represent the two Higgs superfields, $\hat{Q}$, $\hat{u}$ and $\hat{d}$ are the 
quark doublet and  the up-quark and down-quark singlet superfields respectively and
$\hat{L}$ and  $\hat{e}^c$ are the left-handed doublet and right-handed lepton singlet superfields.

The lepton Yukawa matrix $h^e$ 
can be always taken real and diagonal while $h^u$ and $h^d$ contain the KM 
phase. There are four new sources for explicit  $CP$-violating phases, all
arising from parameters that softly break supersymmetry. These are (i) 
the trilinear couplings
\be
\Gamma^u H_2\widetilde{Q}\widetilde{u}^c+\Gamma^d H_1\widetilde{Q}
\widetilde{d}^c+\Gamma^e H_1\widetilde{L}\widetilde{e}^c+{\rm h.c.} \ ,
\ee
where we have defined
\be
\Gamma^{(u,d,e)}\equiv A^{(u,d,e)}\cdot h^{(u,d,e)} 
\ee
and the tildes stand for scalar fields, (ii) the bilinear coupling in the Higgs sector $\mu B H_1 H_2$, (iii) the 
gaugino 
masses $M$ (one for each gauge group), and (iv) the soft scalar masses 
$\widetilde{m}$. 
Two phases may be removed by redefining the phase of the superfield 
$\hat{H}_2$ in such a  way that the phase of $\mu$ is opposite to that of 
$B$. The product $\mu B$ is therefore real. It is also possible to remove 
the phase of the gaugino masses $M$ by an $R$ symmetry transformation. The 
latter leaves all the other supersymmetric couplings invariant, and only 
modifies the trilinear ones, which get multiplied by ${\rm exp}(-\phi_M)$ 
where $\phi_M$ is the phase of $M$. The phases which remain are therefore
\be
\phi_A={\rm arg}(A M)\:\:\:{\rm and}\:\:\:\phi_\mu=-{\rm arg}(B).
\ee
These new phases $\phi_A$ and $\phi_\mu$ may be crucial for the generation 
of the baryon asymmetry. 

In the MSSM, however, there are other possible sources 
of $CP$-violation. In fact, when supersymmetry breaking occurs, as we know 
it must, 
the interactions of the Higgs fields
$H_1$ and $H_2$ with charginos, neutralinos and stops (the superpartners of 
the 
charged, neutral gauge bosons, Higgs bosons and tops, respectively) at the 
one-loop level,
lead to a $CP$-violating contribution to the scalar potential of the form 
\be
V_{CP} = \lambda_1(H_1 H_2)^2 +\lambda_2|H_1|^2 H_1 H_2 +
\lambda_3|H_2|^2 H_1 H_2 + {\rm h.c.}.
\ee
These corrections  may be quite 
large  at finite temperature \cite{cpr1,riottospont}.
One may  write the Higgs fields in
unitary gauge as $H_1=( \varphi_1,0)^T$  and $H_2=(0, 
\varphi_2 e^{i\theta})^T$, 
where $\varphi_1$, $\varphi_2$, $\theta$ are real, and $\theta$ is the 
$CP$-odd phase.
The  coefficients $\lambda_{1,2,3}$  
determine whether the $CP$-odd field $\theta$ is non zero, signalling the 
spontaneous violation of $CP$ in a
similar way to what happens in a generic  two-Higgs model \cite{gu}. If, 
during the evolution of the early universe, 
a variation of the phase $\theta$ is induced,  
this $CP$ breakdown may bias the production of baryons through sphaleron 
processes in 
electroweak baryogenesis models.
The attractive 
feature of this possibility is that, when the universe cools down, the 
finite temperature loop corrections to the $\lambda$-couplings disappear, 
and the phase $\theta$
relaxes to zero. This is turn implies that, unlike for scenarios utilizing
$\phi_A$ and $\phi_\mu$, one need not worry about 
present experimental constraints from the physics of $CP$-violation. 
When the constraints from 
experiment and the strength of the transition are taken into account~\cite{lainespon}, 
the relevant region of the parameter space for spontaneous $CP$-violation in the 
MSSM at finite temperature is rather restricted. Nevertheless, in this 
small region, perturbative estimates need not be reliable, and 
non-perturbatively the region might be slightly larger (or smaller). 

Finally, in scenarios in which the BAU is 
generated from the coherent production of a scalar condensate along a flat 
direction of the supersymmetric extension of the SM, $CP$-violation is present
in the nonrenormalizable superpotentials which lift the flat directions at 
large field values. I will have more to say about this is section (\ref{AD}).

\subsection{Departure from Thermal Equlibrium}

\subsubsection{The Out-of-Equilibrium Decay Scenario}

Scenarios in which the third Sakharov condition is satisfied due to 
the presence of 
superheavy decaying particles in a rapidly expanding universe, 
generically fall under the name 
of out-of-equilibrium decay mechanisms.
The underlying idea is fairly simple. 
If the decay rate $\Gamma_X$ of the superheavy particles $X$ 
at the time they become 
nonrelativistic ({\it i.e.} at the temperature $T\sim M_X$) is much smaller 
than the expansion rate of the universe, then the $X$  
particles cannot decay on the time scale of the expansion and so they 
remain as 
abundant as photons for $T\lsim M_X$. In other words, at some  temperature 
$T > M_X$, the superheavy particles  
$X$ are so weakly interacting that they cannot catch up with the expansion 
of the universe and they decouple from the thermal bath while they are still 
relativistic, so that $n_{X}\sim n_\gamma\sim T^3$ at the time of decoupling. 
Therefore, at temperature $T\simeq M_X$, they populate the universe 
with an abundance which is much larger than the equilibrium one. 
This overbundance is precisely 
the departure from thermal equilibrium needed to produce a final nonvanishing 
baryon asymmetry when the heavy states $X$ undergo $B$ and $CP$-violating
decays.
The out-of-equilibrium condition requires very heavy states: 
$ M_X\gsim (10^{15}-10^{16})\:{\rm GeV}$ and $M_X\gsim  
(10^{10}-10^{16})\:{\rm GeV}$, for gauge and scalar bosons, respectively, 
if these heavy particles decay through renormalizable operators.

\subsubsection{The Electroweak Phase Transition Scenario}
\label{EWPT}
At
temperatures around the electroweak scale, the expansion rate of the universe
in thermal units is small compared to the rate of baryon number violating 
processes. This
means that the equilibrium description of particle phenomena is extremely
accurate at electroweak temperatures. Thus, baryogenesis cannot occur at
such low scales without the aid of phase transitions and the question of 
the order of the electroweak phase transition becomes central. 

If the 
EWPT is second order or a continuous crossover, the associated departure from
equilibrium is insufficient to lead to relevant baryon number production
\cite{krs}.
This means that for EWBG to succeed, we either need the EWPT to be strongly
first order or other methods of destroying thermal equilibrium to be 
present at the phase transition.

Any thermodynamic quantity that undergoes such a 
discontinuous change at a phase transition is referred to as an 
{\it order parameter}, denoted by $\vp$. For a first order transition 
there is an  extremum at $\vp=0$ which becomes separated
from a second local minimum by an energy barrier.
At the critical temperature $T=T_c$ both phases are equally 
favored energetically and at later times the minimum at $\vp \neq 0$ becomes
the global minimum of the theory. 
The dynamics of the phase tansition in this situation
is crucial to most scenarios of electroweak baryogenesis. The essential picture
is that around $T_c$ quantum tunneling
occurs and nucleation of bubbles of the true vacuum 
in the sea of false begins. Initially these bubbles are not large 
enough for their volume energy to overcome the competing surface tension and 
they shrink and disappear. However, at a particular temperature below 
$T_c$, bubbles
just large enough to grow nucleate. These are termed {\it critical} bubbles,
and they expand, eventually filling all of space and completing the transition.
As the bubble walls
pass each point in space, the order
parameter changes rapidly, as do the other fields, and this leads to a
significant departure from thermal equilibrium. Thus, if the phase 
transition is strongly enough first order it is possible to satisfy
the third Sakharov criterion in this way.

The precise evolution of critical bubbles in the electroweak phase transition 
is a crucial factor in determining which regimes of electroweak
baryogenesis are both possible and efficient enough to produce the BAU. 
(See~\cite{G 93,G 94,G&R 94,G 95,B&G 95,G&H 96,B&G 97,GHK 97,Gleiser:1999xd}
for possible obstacles to the standard picture)
I'll discuss this, and the issue of the order of the phase transition, in more detail
in section~\ref{EWBG}.

For the bubble wall scenario to be successful, there is a further 
criterion to be satisfied. As the wall passes a
point in space, the Higgs fields evolve rapidly and the Higgs VEV changes from
$\langle\phi\rangle=0$ in the unbroken phase to
\be
\langle\phi\rangle=v(T_c)
\label{vatTc}
\ee
in the broken phase. Here, $v(T)$ is the value
of the order parameter at the symmetry breaking global minimum of the finite 
temperature effective potential. 
Now, $CP$-violation and the departure from equilibrium occur while the 
Higgs field 
is changing. Afterwards, the point is
in the true vacuum, baryogenesis has ended, and baryon number violation
is exponentially supressed. Since baryogenesis is now over, 
it is
imperative that baryon number violation be negligible at this temperature in
the broken phase, otherwise any baryonic excess generated will be
equilibrated to zero. Such an effect is known as {\it washout} of the 
asymmetry and the criterion for this not to happen may be written as
\be
\frac{v(T_c)}{T_c} \gsim 1 \ .
\label{washout}
\ee
Although there are a number of nontrivial steps
that lead to this simple criterion, (\ref{washout}) is traditionally used 
to ensure that the baryon asymmetry survives after
the wall has passed.
It is necessary that this criterion be satisfied for any electroweak 
baryogenesis scenario to be successful.

\subsubsection{Defect-Mediated Baryogenesis}

A natural way to depart from equilibrium is provided by the dynamics of
topological defects. Topological defects are regions of trapped energy 
density which can
remain after a cosmological phase transition if the topology of the 
vacuum of the theory is nontrivial. Typically, cosmological phase
transitions occur when a symmetry of a particle physics theory
is spontaneously broken. In that case, the cores of the topological 
defects formed are regions in which the symmetry of the unbroken theory
is restored.
If, for example, cosmic strings are produced at the GUT phase 
transition, then the decays of loops of string act as an additional
source of superheavy bosons which undergo baryon number violating
decays. When defects are produced at the TeV scale, 
a detailed analysis of the dynamics of a
network of these objects shows that a significant baryon to entropy
ratio can be generated 
\cite{{DMEWBGii},{DMEWBGv},{DMEWBGiii},{DMEWBGiv},{rob-annei},{rob-anneii}} if 
the electroweak symmetry is  restored around such a higher scale ordinary 
defect \cite{{PD 93},{ANO},{MT 94}}. 
Although a recent analysis has shown that $B$-violation  is highly 
inefficient along nonsuperconducting strings \cite{cemr}, there remain
viable scenarios involving other ordinary defects, superconducting strings 
or defects carrying  baryon number \cite{riottodefect1,riottodefect2}.

\section{Grand-Unified Baryogenesis and Leptogenesis}
\label{GUT}

\subsection{GUT Baryogenesis}
As we have seen, in GUTs baryon number violation is natural
since quarks and leptons lie in the same irreducible representation
of the gauge group. The implementation of the out-of-equilibrium scenario 
in this context goes under the name of GUT baryogenesis. 
The basic assumption   is that the superheavy bosons were as 
abundant as photons at very high temperatures $T\gsim M_X$. This assumption 
is questionable if  the heavy $X$ particles are the gauge or Higgs bosons of 
Grand 
Unification,  because they might have never been in thermal equilibrium in 
the early universe. Indeed, 
 the temperature of the universe might have been  always smaller than  
$M_{{\rm GUT}}$ and, correspondingly, the thermally produced $X$ bosons might 
never have been as abundant as photons, making their role in baryogenesis 
negligible. 
All these considerations depend crucially upon the thermal history of the 
universe  and deserve a closer look. 

The flatness and the horizon problems of the standard big bang
cosmology are elegantly solved if, during the evolution of the early
universe, the energy density happened to be dominated by some form of
vacuum energy, causing comoving scales grow quasi-exponentially 
\cite{guth,new1,new2,reviewinflation}.
The vacuum energy driving this {\it inflation} is generally
assumed to be associated with the potential $V(\phi)$ of  some scalar field 
$\phi$, the {\it inflaton},
which is initially displaced from the minimum of its potential. As a
by-product, quantum fluctuations of the inflaton field may be the seeds
for the generation of structure and the 
fluctuations observed in the cosmic microwave background radiation
\cite{reviewinflation}.
Inflation ended when the potential energy associated with the inflaton
field became smaller than the kinetic energy of the field.  By that
time, any pre-inflationary entropy in the universe had been inflated
away, and the energy of the universe was entirely in the form of
coherent oscillations of the inflaton condensate around the minimum of
its potential.  Now, we know that somehow this low-entropy cold universe
must be transformed into a high-entropy hot universe dominated by
radiation. The process by which the energy of the inflaton field is
transferred from the inflaton field to radiation has been dubbed
{\it reheating}.  
The simplest way to envision this process is if the comoving energy
density in the zero mode of the inflaton decays {\it perturbatively} into 
normal particles,
which then scatter and thermalize to form a thermal background 
\cite{dolgov,abbot}.  It is
usually assumed that the decay width of this process is the same as
the decay width of a free inflaton field.
Of particular interest is a quantity known as the reheat temperature,
denoted as $T_r$. This is calculated by assuming 
an instantaneous conversion of the energy density in the inflaton 
field into radiation when the decay width of the inflaton energy,
$\Gamma_\phi$, is equal to $H$, the expansion rate of the universe. This
yields

\be
T_r= \sqrt{ \Gamma_\phi \mpl }.
\ee
There are very good reasons to suspect that GUT baryogenesis does not occur 
if this is the way reheating happens. First of 
all, the density and temperature fluctuations observed in the present
universe require the inflaton
potential to be extremely flat  -- that is $\Gamma_\phi\ll M_\phi$. This 
means that the couplings of the
inflaton field to the other degrees of freedom  cannot be too large, since 
large couplings would induce
large loop corrections to the inflaton potential, spoiling its
flatness.  As a result, $T_{r}$ is expected to be  smaller than
$10^{14}$ GeV by several orders of magnitude. On the other hand,  the 
unification scale is generally assumed to be around $10^{16}$ GeV,
and $B$-violating  bosons should have masses comparable to this
scale. Furthermore,  even the light $B$-violating Higgs bosons are
expected to have masses larger than the inflaton mass, and thus it would be
kinematically impossible to create them directly in $\phi$ decays, 
$\phi\rightarrow X\overline{X}$.

One might think that the heavy bosons could be created during the stage of 
thermalization. Indeed, particles of mass 
$\sim 10^4$ bigger than the reheating
temperature $T_{r}$ may be created by the thermalized decay products
of the inflaton \cite{ckr1,ckr2,ckr3}.  However, there is one more problem 
associated with GUT baryogenesis in the old theory of reheating, namely the 
problem of relic gravitinos \cite{gravitino}. If one has to invoke 
supersymmetry to preserve the flatness of the inflaton potential, it is 
mandatory to consider the cosmological implications of the gravitino -- a 
spin-(3/2) particle which appears in the  extension of global supersymmetry 
to supergravity. The gravitino is
 the fermionic superpartner of the graviton and has interaction strength 
with the observable sector -- that is the standard model particles and their 
superpartners -- inversely proportional to the Planck mass. The slow gravitino
decay rate leads to a cosmological problem because the  decay 
products of the gravitino destroy $^4$He and D nuclei by 
photodissociation, 
and thus ruin the successful predictions of nucleosynthesis. The requirement
that not too many gravitinos are produced after inflation provides a
stringent constraint on the reheating temperature,   
$T_{r}\lsim (10^{10}-10^{11})\: {\rm GeV}$ 
\cite{gravitino}. Therefore, if $T_{r}\sim M_{{\rm GUT}}$, gravitinos would 
be abundant during nucleosynthesis and destroy the agreement of the 
theory with observations. However, if the initial state after inflation was 
free from gravitinos, the reheating temperature is then too low to 
create superheavy $X$ bosons that eventually decay and produce the baryon 
asymmetry.

The outlook for GUT baryogenesis has brightened with the
realization that reheating may differ significantly from the simple
picture described above \cite{explosive,KT1,KT2,KT3,KT4}.  In the
first stage of reheating, called {\it preheating} \cite{explosive},
nonlinear quantum effects may lead to extremely effective
dissipative dynamics and explosive particle production, even when
single particle decay is kinematically forbidden. In this picture, 
particles can be
produced in a regime of broad parametric resonance, and it is
possible that a significant fraction of the energy stored in the form
of coherent inflaton oscillations at the end of inflation is released
after only a dozen or so oscillation periods of the inflaton. What is
most relevant for us is that preheating may play
an extremely important role in baryogenesis and, in 
particular, for the Grand Unified generation of a baryonic
excess. Indeed, it was shown in \cite{klr,krt} that the baryon
asymmetry can be produced efficiently just after the preheating era,
thus solving many of the problems that GUT baryogenesis had to face in
the old picture of reheating. Interestingly, preheating may also play an important
role in electroweak baryogenesis~\cite{Krauss:1999ng,Garcia-Bellido:1999sv}
in models with very low-scale inflation~\cite{Copeland:2001qw}, although 
I do not have space to discuss this here.

\subsection{Baryogenesis Via Leptogenesis}

Since the linear  combination $B-L$ is left unchanged by sphaleron 
transitions, 
the baryon asymmetry may be generated  from a lepton asymmetry \cite{fy}. 
Indeed, sphaleron transition will
reprocess  any  lepton asymmetry   and convert (a fraction of) it into 
baryon number. This is because $B+L$ must be vanishing  and  the final 
baryon asymmetry results to be 
$B\simeq -L$. 
 Once the lepton number is produced, thermal scatterings  redistribute the 
charges. In the 
 high temperature phase of the SM,  the asymmetries of
baryon number $B$ and of $B-L$ are therefore proportional: 
\cite{ks88} 
\be
     B=\left(\frac{8 n_f+4 n_H}{22 n_f+13 n_H}\right)(B-L),
\ee
where $n_H$ is the
number of Higgs doublets. In the SM as well as in its
unified extension based on the group $SU(5)$, $B-L$ is conserved and 
no asymmetry in $B-L$ can be generated.
However, adding  right-handed Majorana neutrinos
to the SM breaks $B-L$ and the primordial lepton asymmetry may be  generated 
by the out-of-equilibrium
decay of heavy right-handed Majorana neutrinos $N_L^c$ (in the supersymmetric 
version,  heavy scalar neutrino decays are also  relevant for leptogenesis).
This simple extension of the SM can be
embedded into GUTs with gauge groups containing
$SO(10)$. Heavy right-handed Majorana neutrinos can also
explain the smallness of the light neutrino masses via the see-saw
mechanism \cite{pati2,goran,gelmann}.

The relevant  couplings are those  between the right-handed neutrino, the 
Higgs doublet $\Phi$ and the lepton doublet $\ell_L$
\be
{\cal L}=
\overline{\ell_L}\:\Phi\:h_\nu\:N_L^c +\frac{1}{2}\overline{N_L^c}\:M\:N_L^c+
{\rm h.c.}
\ee
The vacuum expectation value of the Higgs field $\langle \Phi\rangle$ 
generates neutrino Dirac masses $m_D=h_\nu \langle \Phi\rangle$, which are 
assumed to be much smaller than the Majorana masses $M$. When the  Majorana 
right-handed neutrinos decay into leptons and Higgs scalars, they violate 
the lepton number since right-handed neutrino fermionic lines do not have 
any preferred arrow
\begin{eqnarray}
N_L^c &\rightarrow & \overline{\Phi}+\ell,\nonumber\\
N_L^c &\rightarrow &\Phi+\overline{\ell}.
\end{eqnarray}
The interference between the tree-level decay amplitude and the absorptive 
part of the one-loop vertex  leads to a lepton  asymmetry of the right order 
of magnitude to explain the observed baryon asymmetry and has been discussed 
extensively in the literature \cite{fy,luty1,luty,va}. Recently, much attention
has been paid to the effects of finite temperature on the $CP$-violation 
\cite{covitem}, and  
the contribution to the  lepton asymmetry  generated by the tree-level graph 
and the absorptive part of the one-loop self-energy.
In particular,  it has been observed that $CP$-violation may be 
considerably enhanced if two heavy right-handed  neutrinos are nearly 
degenerate in mass \cite{reviewlep}.
It is also important to remember   that  large lepton number violation at 
intermediate temperatures may potentially dissipate away the baryon number 
in combination with the sphaleron transitions. Indeed,  $\Delta L=2$ 
interactions of the form
\be
\label{zzz}
\frac{m_\nu}{\langle \Phi\rangle^2}\ell_L\ell_L\Phi\Phi +{\rm h.c.},
\ee
where $m_\nu$ is the mass of the light left-handed neutrino, are generated 
through the exchange of heavy right-handed neutrinos. The rate of lepton 
number violation induced by this interaction is therefore 
$\Gamma_L\sim (m_\nu^2/\langle \Phi\rangle^4) T^3$. The requirement of 
harmless letpon number violation, $\Gamma_L\lsim H$ imposes an interesting 
bound on the neutrino mass
\be
m_{\nu}\lsim 4\:{\rm eV}\left(\frac{T_X}{10^{10}\:{\rm GeV}}\right)^{-1/2},
\ee
where $T_X\equiv {\rm Min}\left\{T_{B-L},10^{12}\:{\rm GeV}\right\}$, 
$T_{B-L}$ is the temperature at which the $B-L$ number production takes place, 
and $\sim 10^{12}$ GeV is the temperature at which sphaleron transitions 
enter in equilibrium. One can also reverse the argument and  study 
leptogenesis assuming a similar pattern of mixings and masses for leptons 
and quarks, as suggested by $SO(10)$ unification \cite{buch}. This implies 
that $B-L$ is broken at the unification scale $\sim 10^{16}$ GeV, if 
$m_{\nu_{\mu}} \sim 3\times  10^{-3}$ eV as preferred by the MSW explanation 
of the solar neutrino deficit \cite{wolf,ms}. 

\section{Electroweak Baryogenesis}
\label{EWBG}
Scenarios in which anomalous baryon number violation in the electroweak 
theory implements the first Sakharov criterion are refered to as
{\it electroweak baryogenesis}. Typically these scenarios require a strongly 
first order phase transition, either in the minimal standard model, or in a 
modest extension. Another possibility is that the out of equilibrium 
requirement is achieved through TeV scale topological defects, also present in
many extensions of the electroweak theory. Finally, a source of $CP$-violation
beyond that in the CKM matrix is usually involed.

Historically, the ways in which baryons may be produced as a bubble wall, or
phase boundary, sweeps through space, have been separated into two
categories. 

\begin{enumerate}
\item {\it local baryogenesis}: baryons are produced when the baryon number 
violating processes and $CP$-violating processes occur together near 
the bubble walls. 
\item {\it nonlocal baryogenesis}: particles undergo $CP$-violating
interactions with the bubble wall and carry an asymmetry in a quantum number 
other than baryon number into the unbroken phase region away from the wall.
Baryons are then produced as baryon number violating processes convert the
existing asymmetry into one in baryon number.
\end{enumerate}

In general, both local and nonlocal baryogenesis will occur and the BAU will
be the sum of that generated by the two processes.
However, if the speed of the wall is greater than the sound speed in the
plasma, then local baryogenesis dominates~\cite{LRT 97}. In other cases, 
nonlocal baryogenesis is usually more efficient and I will focus on that
here.

Nonlocal baryogenesis typically involves the interaction of the 
bubble wall with the various fermionic species in the unbroken phase.
The main picture is that as a result of $CP$-violation in the bubble wall, 
particles with opposite chirality interact differently with the wall,
resulting in a net injected chiral flux. This flux
thermalizes and diffuses into the unbroken phase where it is
converted to baryons.
In this section, for definiteness when describing these 
effects, I assume that the $CP$-violation arises because of a two-Higgs
doublet structure. 

The chiral asymmetry which is converted to an asymmetry in baryon
number is carried by both quarks and leptons. However, the
Yukawa couplings of the top quark and the $\tau$-lepton are larger
than those of the other quarks and leptons respectively. Therefore, it is
reasonable to expect that the main contribution to the injected asymmetry 
comes from these particles and to neglect the effects of the 
other particles.   

When considering nonlocal baryogenesis it is convenient to write the 
equation for the rate of production of baryons in the form \cite{JPT2 94}
\begin{equation}
\frac{d n_B}{dt} = -\frac{n_f\Gamma_{sp}(T)}{2T}\sum_i\mu_i \ ,
\label{nonlocal B}
\end{equation}
where the rate per unit volume for electroweak sphaleron transitions 
is given by~(\ref{unbrokenrate}).
Here, $n_f$ is again the 
number of families and $\mu_i$ is the chemical potential for left 
handed particles of species $i$. The crucial question in applying this 
equation is an accurate evaluation of the chemical potentials that bias 
baryon number production.

There are typically two distinct calculational regimes
that are appropriate for the treatment of nonlocal effects in electroweak
baryogenesis. Which regime is appropriate depends on the particular fermionic
species under consideration.

\subsubsection{The Thin Wall Regime}

The thin wall regime \cite{{CKN2i},{CKN2ii},{JPT2 94}} applies if
the mean free path $\ell$ of the fermions being considered is much greater
than the thickness $L_w$ of the wall, i.e. if
\be
\frac{L_w}{\ell}\lsim 1 \ .
\ee 
In this case we may neglect scattering effects and
treat the fermions as free particles in their interactions with the wall.   

In this regime, in the rest frame of the bubble wall, particles see a sharp 
potential barrier 
and undergo $CP$-violating interactions with the wall due to the gradient in the 
$CP$ odd Higgs phase. 
As a consequence of $CP$-violation, there will be asymmetric reflection
and transmission of particles, thus generating an injected current
into the unbroken phase in front of the bubble wall. As a consequence of this
injected current, asymmetries in certain quantum numbers will diffuse
both behind and in front of the wall due to particle
interactions and decays \cite{{CKN2i},{CKN2ii},{JPT2 94}}. In particular, the
asymmetric reflection and transmission of left and right handed
particles will lead to a net injected chiral flux from the wall.
However, there is a qualitative difference between the diffusion
occurring in the interior and exterior of the bubble.        
 
Exterior to the bubble the electroweak symmetry is restored and
weak sphaleron transitions are unsuppressed. This means that the
chiral asymmetry carried into this region by transport of the injected
particles may be converted to an asymmetry in baryon number by
sphaleron effects. In contrast, particles injected into the phase of
broken symmetry interior to the bubble may diffuse only by baryon number 
conserving decays since the electroweak sphaleron rate is exponentially 
suppressed in this region. Hence, we concentrate only on those particles
injected into the unbroken phase.

The net baryon to entropy ratio which results via nonlocal
baryogenesis in the case of thin walls has been calculated in several
different analyses \cite{{CKN2i},{CKN2ii}} and \cite{JPT2 94}. 
In the following I will give a brief outline of the logic of the calculation, 
following \cite{JPT2 94}. 
The baryon
density produced is given by~(\ref{nonlocal B}) in
terms of the chemical potentials $\mu_i$ for left handed particles. 
These chemical potentials are a consequence of the asymmetric reflection and
transmission off the walls and the resulting chiral particle asymmetry.       
Baryon number
violation is driven by the chemical potentials for left handed leptons
or quarks. To be concrete, focus on leptons \cite{JPT2 94} (for  
quarks see e.g. \cite{CKN2i}). If there is local thermal
equilibrium in front of the bubble walls then the
chemical potentials $\mu_i$ of particle species $i$ are related to
their number densities $n_i$ by
\begin{equation}  
n_i = {{T^2} \over {12}} k_i \mu_i \ ,
\end{equation}
where $k_i$ is a statistical factor which equals $1$ for fermions and $2$ 
for bosons. In deriving this expression, it is important \cite{JPT1 94}
to correctly impose the 
constraints on quantities which are conserved in the region in front of and 
on the wall.

Using the above considerations, the chemical potential $\mu_L$ for left 
handed leptons can be related to the left handed lepton number densities 
$L_L$. These are in turn determined by particle transport. The source term 
in the diffusion equation is the flux $J_0$ resulting from the 
asymmetric reflection and transmission of left and right handed leptons 
off the bubble wall.

For simplicity assume a planar wall. 
If $\vert p_z \vert$ is the momentum of the lepton perpendicular to 
the wall (in the wall frame), the analytic approximation used in
\cite{JPT2 94} allows the asymmetric reflection coefficients for 
lepton scattering to be calculated in the range
\be
m_l < \vert p_z \vert < m_h \sim {1 \over L_w} \ ,
\label{range}
\ee
where $m_l$ and $m_h$ are the lepton and Higgs masses, respectively, 
and results in
\begin{equation}
{\cal R}_{L \rightarrow R} - {\cal R}_{R \rightarrow L} \simeq 2 
\Delta \theta_{CP} {{m_l^2} \over {m_h \vert p_z \vert}} \ .
\label{reflection}
\end{equation}
The corresponding flux of left handed leptons is
\begin{equation}
J_0 \simeq {{v_w m_l^2 m_h \Delta \theta_{CP}} 
\over {4 \pi^2}} \ ,
\end{equation}
where $v_w$ is the velocity of the bubble wall. 
Note that in order for 
the momentum interval in~(\ref{range}) to be non-vanishing, the condition 
$m_l L_w < 1$ needs to be satisfied.

The injected current from a bubble wall will lead to a ``diffusion tail" of
particles in front of the moving wall. In the approximation in which  the
persistence length of the injected current is much larger than
the wall thickness we may to a good approximation model it as a delta 
function source and search for a steady state solution. In 
addition, assume that the 
decay time of leptons is much longer than the time it takes for a
wall to pass so that we may neglect decays. Then the diffusion equation
for a single particle species is easily solved by
\begin{equation}
L_L(z) = \left\{ \begin{array}{ll}
          J_0 {{\xi_L} \over {D_L}} e^{- \lambda_D z}  & \ \ \ \ z>0 \\
          0 & \ \ \ \ z<0
         \end{array} \right. \ ,
\end{equation}
with the diffusion root 
\be
\lambda_D = \frac{v_w}{D_L} \ ,
\ee
where $D_L$ is the diffusion constant for leptons, and $\xi^L$ 
that is called the {\it persistence length} of the current in front of 
the bubble wall.

Note that in this approximation the injected current does not generate
any perturbation behind the wall. This is true
provided $\xi^L \gg L_w$ is satisfied. If this  
inequality is not true,  the problem becomes significantly more complex
\cite{JPT2 94}. 

In the massless approximation the chemical potential $\mu_L$ can be
related to $L_L$ by 
\begin{equation}
\mu_L = { 6 \over {T^2}} L_L
\end{equation}
(for details see \cite{JPT2 94}).
Inserting the sphaleron rate and the above results for the chemical potential 
$\mu$ into~(\ref{nonlocal B}), and using $1/D_L \simeq 8 \alpha_W^2 T$, 
the final baryon to entropy ratio becomes
\begin{equation}
\frac{n_B}{s} \sim 0.2\, \alpha_W^2 (g^*)^{-1} \kappa
\Delta \theta_{CP} \frac{1}{v_w} \left(\frac{m_l}{T}\right)^2 
\frac{m_h}{T} \frac{\xi^L}{D_L} \ .
\end{equation}
where I have assumed that sphalerons do not equilibrate in the diffusion tail.

Now consider the effects of top quarks scattering off the
advancing wall \cite{{CKN2i},{CKN2ii}}.  Several effects tend to decrease
the contribution of the top quarks relative to that of tau leptons.
Firstly, for typical wall thicknesses the thin wall approximation does not
hold for top quarks. This is because top quarks are much more strongly
interacting than leptons and so have a much shorter mean free path. 
An important effect is that the diffusion tail is cut
off in front of the wall by {\it strong sphalerons} \cite{mz,GS94}. 
There is an anomaly in the quark axial vector current
in QCD. This leads to chirality non-conserving processes at high 
temperatures. These processes are relevant for nonlocal baryogenesis 
since it is the chirality
of the injected current that is important in that scenario. In an analogous
expression to that for weak sphalerons, we may write the rate per unit volume 
of chirality violating processes due to strong sphalerons in the unbroken 
phase as
\be
\Gamma_s = \kappa_s(\alpha_s T)^4 \ ,
\label{strongsphaleron}
\ee
where $\kappa_s$ is a dimensionless constant \cite{GM 97}. Note that the
uncertainties in $\kappa_s$ are precisely the same as those in $\kappa$
defined in (\ref{unbrokenrate}). As such, $\kappa_s$ could easily be
proportional to $\alpha_s$, in analogy with
(\ref{unbrokenrate}), perhaps with a logarithmic correction.
These chirality-changing processes damp the effect of
the injected chiral flux and effectively cut off the diffusion tail in front
of the advancing bubble wall. Second, the diffusion
length for top quarks is intrinsically smaller than that for tau leptons, 
thus reducing the volume in which
baryogenesis takes place.  Although there are also enhancement factors, e.g.
the ratio of the squares of the masses $m_t^2/m_{\tau}^2$, it seems
that leptons provide the dominant contribution to nonlocal 
baryogenesis.

\subsubsection{The Thick Wall Regime}

Now let's turn to the thick wall, or adiabatic, regime. This is relevant
if the mean free path of the particles
being considered is smaller than the width of the wall, $\ell\lsim L_w$. When 
the walls are 
thick, most 
interactions within the wall will be almost in thermal equilibrium. The
equilibrium is not exact because some interactions, in particular
baryon number violation, take place on a time scale much slower than the
rate of passage of the bubble wall. These slowly-varying quantities are
best treated by the method of chemical potentials. 
The basic idea is  to produce asymmetries in some local charges which are 
(approximately) conserved by the interactions inside the bubble walls, where 
local departure from thermal equilibrium is attained. These local charges 
will then diffuse into the unbroken phase where  baryon number violation is 
active thanks to the unsuppressed sphaleron transitions.  The latter convert 
the asymmetries into a baryon asymmetry. Therefore, one has to {\it i)} 
identify  
 those  charges which are
approximately conserved in the symmetric phase, so that they
can efficiently diffuse in front of the bubble where baryon number
violation is fast, {\it ii)}  compute the $CP$-violating  currents  of the 
plasma induced inside the bubble wall and {\it iii)}   
solve a set of coupled differential diffusion equations for the local 
particle densities, including the $CP$-violating  sources. 

There are a number of possible
$CP$-violating sources.
To be definite, consider the example where $CP$-violation is due to a
$CP$ odd phase $\theta$ in the two-Higgs doublet model \cite{CKN4i}. This goes 
generically under the name of spontaneous baryogenesis. 
In order to explicitly see how $\theta$ couples to the 
fermionic sector of the theory (to produce baryons) we may remove the 
$\theta$-dependence of
the Yukawa couplings arising from the Higgs terms. This is done by
performing an anomaly-free hypercharge rotation on the fermions
\cite{CKN4i}, inducing a term in the Lagrangian density of the form $
{\cal L}_{CP} \propto \partial_{\mu}\theta J^\mu_Y$, 
where $J^\mu_Y$ is the  fermionic part of the hypercharge current. 
Therefore, a nonvanishing $\dot{\theta}$ provides a preferential direction for 
the production of quarks and leptons; in essence a chemical potential 
$\mu_B\propto \dot{\theta}$ (in the rest frame of the thermal bath)
for baryon number. Of course, strictly speaking, this is not a chemical 
potential since it 
arises dynamically rather than to impose a constraint. For this reason,
the quantity $\mu_B$ is sometimes referred to as a ``charge potential''.
A more complete field theoretic approach to the computation of particle 
currents on a space-time dependent and $CP$-violating Higgs background was 
provided in \cite{CPR95,CPR96} where it was shown that  fermionic currents 
arise at one loop, and are proportional to $(h v(T_c)/\pi T)^2 \dot{\theta}$, 
where $h$ is the appropriate Yukawa coupling. The relevant baryon number 
produced by spontaneous baryogenesis is then
calculated by integrating 
\be
\frac{dn_B}{dt} = -9\frac{\Gamma_{sp}(T)}{T} \mu_B \ .
\ee
Initially, spontaneous baryogenesis was considered as an
example of local baryogenesis. However, it has become clear that diffusion
effects can lead to an appreciable enhancement of the baryon asymmetry 
produced by this mechanism \cite{CKN 94,cprdif1}.

An alternative way of generating a source for the diffusion equation was
suggested by Joyce et al. \cite{JPT3 94}. The essential idea is that there 
exists 
a purely classical chiral force that acts on particles when there is a $CP$ 
violating field on the bubble wall.

To summarize, the effects of transport in the plasma 
external to the expanding bubble allow baryon violating transitions in the 
unbroken phase to transform charge  asymmetries produced on the wall into a 
baryon asymmetry. This means that in the case of nonlocal baryogenesis we 
do not need to rely on baryon number violating processes occurring in the
region where the Higgs fields are changing.
The diffusion equation approach to the problem of nonlocal baryogenesis has 
been very successful. In the thin wall case, it is a valid approximation
to assume that the source for the diffusion equation is essentially a
$\delta$-function on the wall. This is because one may ignore the effects
of particle scattering in this picture. However, in the case of thick
walls, significant particle scattering occurs and as a result it is necessary
to consider sources for the diffusion equations that extend over the wall.

\subsection{Electroweak Baryogenesis in the MSSM}

As I have mentioned, the most promising and  
well-motivated framework incorporating $CP$-violation beyond the SM and an
enhanced electroweak phase transition seems to be supersymmetry.
Electroweak  
baryogenesis in the framework of the  
MSSM has  attracted much attention in the past years, with 
particular emphasis on the strength of the phase 
transition~\cite{early1,early2,early3,early4} and  
the mechanism of baryon number generation 
\cite{nelson,noi,riottospont,cpt,cpt1,ck}.

The behavior of the electroweak phase transition in the minimal 
supersymmetric standard model is dependent on the mass of the lightest Higgs
particle, and the mass of the top squark.
Recent  analytical \cite{r1,r2,r3,bod1,losada1,losada2,r4,r5,losada3} and  
lattice  
computations  \cite{r6,r7,r8,r9} have  revealed   that the phase transition 
can be sufficiently strongly  
first order  in the presence of a top squark
lighter than the top quark.
In order to naturally
suppress contributions to the $\rho$-parameter, and hence
preserve a good agreement with precision electroweak measurements,
the top squark should be mainly right handed. This can be achieved if the left
handed stop soft supersymmetry breaking mass $m_Q$
is much larger than $M_Z$.
For moderate mixing, the lightest stop mass is then approximately given by
\be
\label{app}
\mstop^2  \simeq m_U^2  + m_t^2(\phi) \left( 1  -
\frac{\left|\widetilde{A}_t\right|^2}{m_Q^2}
\right) \ ,
\ee
where $\widetilde{A}_t = A_t - \mu^{*}/\tan\beta$ is the
particular combination appearing in the off-diagonal terms of
the left-right stop squared mass matrix and $m_U$ is
the soft supersymmetry breaking mass parameter
of the right handed stop.  
 
The preservation of the baryon number asymmetry requires  the
order parameter $\langle \phi(T_c)\rangle /T_c$ to be larger than one, see 
Eq. (\ref{washout}). This quantity is bounded from above
\be
\frac{\langle \phi(T_c)\rangle}{T_c} < 
\left(\frac{\langle \phi(T_c)\rangle}{T_c}\right)_{\rm SM}
+ \frac{2 \; m_t^3  \left(1 -
\widetilde{A}_t^2/m_Q^2\right)^{3/2}}{ \pi \; v \; m_h^2}\ ,
\label{totalE}
\ee
where $m_t = \overline{m}_t(m_t)$ is the on-shell running top
quark
mass in the $\overline{{\rm MS}}$ scheme, $m_h$ is the lightest Higgs boson mass and $v=\sqrt{v_1^2+v_2^2}$ is the vev  of the Higgs fields.
The first term on the right hand side of
expression ~(\ref{totalE}) is the Standard Model contribution
\be
\left(\frac{\langle \phi(T_c)\rangle}{T_c}\right)_{\rm SM}
\simeq \left(\frac{40}{m_h[{\rm GeV}]}\right)^2,
\ee
and the second term is
the contribution that would be obtained if the right handed
stop plasma mass vanished at the critical
temperature (see Eq.~(\ref{plasm})). The difference between the SM and the 
MSSM is that light stops may give a large contributions to the effective 
potential in the MSSM and therefore overcome the Standard Model
constraints.
The stop contribution strongly depends
on the value of $m_U^2$, which must be small in magnitude, and
negative, in order to induce a sufficiently strong first order phase
transition. Indeed, large stop contributions
are always associated with small values of the right handed stop
plasma mass
\begin{equation}
m^{\rm eff}_{\;\widetilde{t}} = -\widetilde{m}_U^2 + \Pi_R(T) \ ,
\label{plasm}
\end{equation}
where $\widetilde{m}_U^2 = - m_U^2$, and
\be
\Pi_R(T) \simeq 4 g_3^2
T^2/9+h_t^2/6[2-\widetilde{A}_t^2/m_Q^2]T^2
\ee
is the finite temperature self-energy contribution to the right-handed
squarks. Moreover, the
trilinear mass term, $\widetilde{A}_t$,
must satisfy $\widetilde{A}_t^2 \ll m_Q^2$
in order to avoid the suppression of  the stop contribution
to $\langle \phi(T_c)\rangle/T_c$.
Note that, although large values of $\widetilde{m}_U$, of order of the critical
temperature, are useful to get a strongly first order phase transition, 
$\widetilde{m}_U$ 
is bounded from above to avoid the appearance of dangerous charge and color 
breaking minima \cite{r5}. 

As is clear from (\ref{totalE}), in order to obtain values of $\langle 
\phi(T_c)\rangle/T_c$ larger than one,
the Higgs mass must take small values, close to the present
experimental bound.  
For small mixing and large $m_A$, the one-loop Higgs mass has a very simple 
form
\begin{equation}
m_h^2 = M_Z^2 \cos^2 2\beta + \frac{3}{4\pi^2}
\frac{\overline{m}_t^4}{v^2}
\log\left(\frac{m_{\widetilde{t}}^2 m_{\widetilde{T}}^2}
{\overline{m}_t^4}\right)\left[1
+ {\cal{O}}\left(\frac{\widetilde{A}_t^2}{m_Q^2}\right)
\right],
\end{equation}
where $m_{\widetilde{T}}^2 \simeq m_Q^2 + m_t^2$. Hence,  small values of 
$\tan\beta$ are preferred. The
larger the left handed stop mass, the closer to one $\tan\beta$
must be. This implies that
the left handed stop effects are likely to decouple at the critical 
temperature, and hence that $m_Q$ mainly affects the baryon asymmetry through
the resulting Higgs mass.  A detailed analysis~\cite{Carena:2002ss}, including all dominant 
two-loop finite temperature corrections to the Higgs effective potential and  
the non-trivial effects arising from mixing in the stop sector and taking into account  the 
experimental bounds as well as the requirement of avoiding dangerous color 
breaking minima, concludes that the lightest Higgs should be lighter than 
about $120$ GeV, while the stop mass must be smaller than, or of order of, the top quark 
mass. This lower bound  has been essentially confirmed  by lattice 
simulations \cite{r9}, providing a motivation for the search 
for Higgs and stop particles.

As we have seen, the  MSSM contains additional sources  
of $CP$-violation  besides the CKM matrix phase. 
These new phases are essential  for the generation of the baryon number 
since  large  
$CP$-violating  sources may be  locally induced by the passage of the bubble 
wall separating the broken from the unbroken phase during the electroweak 
phase transition. 
The new phases   
appear    in the soft supersymmetry breaking  
parameters associated with the stop mixing angle and the gaugino and  
neutralino mass matrices. However, large values of the stop mixing angle
are strongly restricted in order to preserve a
sufficiently strong first order electroweak phase transition. 
Therefore, an acceptable baryon asymmetry 
may only be generated from the stop sector through a delicate balance 
between the values
of the different soft supersymmetry breaking parameters contributing
to the stop mixing parameter, and their associated $CP$-violating 
phases \cite{noi}. As a result, the contribution to the final baryon 
asymmetry from the stop sector turns out to be negligible.   On the other 
hand, charginos and neutralinos may be responsible for the observed baryon 
asymmetry if a large $CP$-violating source for the higgsino number is 
induced in the bubble wall. This can happen for sufficiently large values 
of the phase of the parameter $\mu$ \cite{noi,ck}. 

\section{Affleck-Dine Baryogenesis}
\label{AD}
Now let's turn to another baryogenesis scenario that has received a 
lot of attention. This mechanism was introduced by Affleck and Dine (AD) 
\cite{affleckdine} and involves the cosmological evolution of scalar fields 
carrying baryonic charge. As we shall see, these scenarios are naturally 
implemented in the context of supersymmetric models. Let us quickly describe 
this before presenting a detailed analysis.

Consider a colorless, electrically neutral combination of quark and
lepton fields. In a supersymmetric theory this object has a scalar 
superpartner, $\chi$, composed of the corresponding squark ${\tilde q}$ 
and slepton ${\tilde l}$ fields. 

Now, as I mentioned earlier, an important feature of supersymmetric field 
theories is the existence of ``flat directions" in field space, on which 
the scalar potential vanishes. Consider the case where some component of 
the field $\chi$ lies along a flat direction. By this I mean that there exist
directions in the superpotential along which the relevant components of 
$\chi$ can be considered as a free massless field.
At the level of renormalizable terms, flat directions are generic, but
supersymmetry breaking and nonrenormalizable operators lift the flat 
directions and sets the scale for  their potential.

During inflation it is likely that the $\chi$ field is displaced from the 
position $\langle\chi\rangle=0$, establishing the initial conditions for 
the subsequent evolution of the field. An important role is played at this 
stage by baryon number violating operators in the potential $V(\chi)$, which 
determine the initial phase of the field. When the Hubble rate becomes of the 
order of the curvature of the potential $\sim m_{3/2}$, the condensate starts 
oscillating around its present minimum. At this time, $B$-violating terms in 
the potential are of comparable importance to the mass term, thereby 
imparting a substantial baryon number to the condensate. After this 
time, the baryon number violating operators are negligible so that, when the 
baryonic charge of $\chi$ 
is transferred to fermions through decays, the  net baryon number of the 
universe is  preserved by the subsequent cosmological evolution. 

In this section I will focus on the most recent developments about AD baryogenesis, referring the reader to \cite{AD 92}  for more details about the scenario as envisaged originally in \cite{affleckdine}. 

\subsection{Supersymmetric Implementations}

The most recent implementations of the Affleck-Dine scenario have been in
the context of the minimal supersymmetric standard model \cite{Lisa1,Lisa2}.
In models like the MSSM, with large numbers of fields, flat directions occur
because of accidental degeneracies in field space. Although a flat direction is
parameterized by a chiral superfield, I shall here focus on just the scalar
component.
In general, flat directions carry a global $U(1)$ quantum number. I will
be interested in those directions in the MSSM that carry $B-L$, since the 
Affleck-Dine condensate will decay above the weak scale and we must avoid 
erasure of any asymmetry through rapidly occurring sphaleron processes. 
Although there are many such directions in the MSSM \cite{Lisa2}, it is
sufficient to treat a single one as typical of the effect. As mentioned 
above, flat directions can be lifted by supersymmetry breaking and 
nonrenormalizable effects in the superpotential. An important obeservation is 
that inflation, by definition, breaks global supersymmetry because of a nonvanishing
cosmological constant (the false vacuum energy density of the
inflaton). In supergravity theories, supersymmetry breaking is
transmitted by gravitational interactions and the supersymmetry breaking  mass 
squared is naturally $C H^2$, where $C={\cal O}(1)$ 
\cite{coughlan,dinefisch}. To illustrate this effect, consider a term in the
K\"{a}hler potential of the form 
\begin{equation}
\label{ch}
\delta K=C\:\int\: d^4\:\theta \frac{1}{\mpl^2}\chi^\dagger\chi\phi^\dagger 
\phi,
\end{equation} 
where $\phi$ is the field which dominates the energy density $\rho$ of the 
universe, that is $\rho\simeq \langle 
\int d^4\theta \phi^\dagger \phi\rangle$. During inflation, $\phi$ is 
identified with the inflaton field and $\rho=V(\phi)=3 H^2\mpl^2$. The 
term (\ref{ch}) therefore provides an effective $\chi$ mass 
$m_{\chi}=-3C H^2$. 
This example can easily be generalized, and the resulting 
potential of the AD flat direction during inflation is of the form  
\begin{equation}
V(\chi)=H^2 \mpl^2 f(\chi/\mpl)+H \mpl^3 g(\chi^n/\mpl^n),
\end{equation}
where $f$ and $g$ are some functions, and the second term is the generalized 
$A$-term coming from nonrenormalizable operators in the superpotential. 

Crucial in assessing the possibility of baryogenesis is the sign of the 
induced mass squared at $\chi=0$.  If the sign of the induced mass squared 
is negative, a large expecation value for a flat direction may develop, which 
is set by the balance with the nonrenormalizable terms in the 
superpotential.  In such a case the salient features of the post-inflationary 
evolution are as follows \cite{Lisa2}:

{\it i)} Suppose that during inflation the Hubble rate is $H_I\gg m_{3/2}$ 
and the  potential is of the form
\begin{equation}
\label{aterm}
V(\chi)=
\left(m_{3/2}-|C| H_I^2\right)|\chi|^2+\left[\frac{\lambda(a H_I+A m_{3/2})
\chi^n}{n M^n}+{\rm h.c.}\right]+|\lambda|^2\frac{|\chi|^{2n-2}}{M^{2n-6}},
\end{equation}
where  $a$ is a constant of order unity, and $M$ is some large mass scale such 
as the GUT or the Planck scale. The AD field evolves exponentially to the 
minimum of the potential, 
$\langle |\chi|\rangle\sim \left(H_I M^{n-3}/\lambda\right)^{1/n-2}$. The 
$A$-term in (\ref{aterm}) violates
the $U(1)$ carried by $\chi$, and the potential in the angular direction is 
proportional to
$\cos(\theta_a+\theta_\lambda+n\theta)$ where $\chi=|\chi|e^{i\theta}$, and
$\theta_a$ and $\theta_{\lambda}$ are the respective phases of $a$ and 
$\lambda$. 
This creates  $n$ discrete minima for the phase of $\chi$. During inflation 
the field quickly settles into one of these minima, establishing the initial 
phase of the field.

{\it ii)} Subsequent to inflation, the minimum of the potential is time 
dependent since $H$ changes with time and the AD field oscillates near the 
time dependent minimum $\langle |\chi|\rangle(t)$ with decreasing amplitiude. 

{\it iii)} When $H\sim m_{3/2}$ the sign of the mass squared becomes positive 
and the field begins to oscillate about $\chi=0$ with frequency $m_{3/2}$ and
amplitude $\langle |\chi|\rangle(t\sim m_{3/2}^{-1})$.
At this time the $B$-violating $A$-terms are as important as the mass term, 
and there is no sense in which the baryon number is conserved.  When the 
field begins to oscillate freely, a large fractional baryon number is generated
during the initial spiral  motion of the field. The important role of $CP$ 
violation is also dictated by the $A$-terms, since the angular term 
$\cos(\theta_A+\theta_\lambda+n\theta)$ becomes important at this stage,  
and a nonzero $\dot{\theta}$ is generated if $\theta_a\neq \theta_A$. This 
is of course necessary since the baryon number is given by 
$n_b=2|\chi|^2\dot{\theta}$. 
When the condensate decays, the baryon asymmetry is transferred to fermions. 

{\it iv)} In the case in which  the inflaton decays when $H<m_{3/2}$ 
(consistently with the requirement that not too many gravitinos are 
produced at reheating), the resulting baryon to entropy ratio is
\begin{equation}
B\simeq \frac{n_b}{n_\chi}\frac{T_r}{m_{3/2}}\frac{\rho_\chi}{\rho_\phi},
\end{equation}
where $n_\chi=\rho_\chi/m_{3/2}$, $\rho_\phi\sim m_{3/2}^2 \mpl^2$ is the 
energy density of the inflaton field when $H\sim m_{3/2}$, and $T_r$ is the 
reheating temperature. For example $n=4$ gives \cite{Lisa2}
\begin{equation}
B\simeq 10^{-10} \left(\frac{T_r}{10^{6}\:{\rm GeV}}\right)
\left(\frac{10^{-3}\:M}{\lambda \mpl}\right).
\end{equation}  

The MSSM contains many combinations of fields for constructing flat 
directions carrying a nonvanshing $B-L$ \cite{Lisa2}. Particularly appealing  
directions are the ones  which carry $B-L$ and can be lifted at $n=4$ level. 
They are   the  $\chi^2=LH_2$ directions. The nonrenormalizable operator is 
then
\begin{equation}
W=\frac{\lambda}{M}\left(L H_2\right)^2,
\end{equation}
which may be present directly at the Planck scale, or could be generated, 
as in $SO(10)$ GUTs, by integrating out right-handed neutrino superfields 
which are heavy standard model singlets.   

\subsection{Initial Conditions for Viable AD Baryogenesis}

From the above picture it is clear that a successful AD baryogenesis 
mechanism is achieved  if the   effective potential during inflation contains 
a negative effective mass term and nonrenormalizable terms that lift the 
flat directions of the potential.   

With a minimal K\"{a}hler potential only, the effective mass squared 
$m^2\sim H_I^2$ during inflation is positive, $\chi=0$ is stable, and the 
large 
expectation values required for baryogenesis do not result. Quantum de Sitter 
fluctuations do excite the field with $\langle \chi^2\rangle\sim H_I^2$, but 
with a correlation length of the order of $H_I^{-1}$. Any resulting baryon 
asymmetry then averages to zero over the present universe. In addition, the 
relative magnitude of the $B$ violating $A$-terms in the potential is small 
for $H_I\ll M$. 

A negative sign for the effective mass squared is possible if one either
considers general K\"{a}hler potentials, or in the case in which inflation is 
driven by  a Fayet-Iliopoulos
$D$ term, which  preserves the flat directions of global supersymmetry,
and in particular keeps the inflaton potential flat, making $D$-term 
inflation particularly attractive \cite{reviewinflation}. If the AD field 
carries a charge of the appropriate sign under the $U(1)$ whose 
Fayet-Iliopoulos term is responsible for inflation, a nonvanishing vacuum 
expectation value may develop \cite{casasgelmini}. Notice that, if the 
AD field is neutral under the $U(1)$, it may nonetheless acquire a negative 
mass squared after the end of $D$-term inflation through nonminimal 
K\"{a}hler potentials relating the AD field to the fields involved in the 
inflationary scenario \cite{kmr}. 

Another logical possibility is that the effective mass squared during 
inflation is positive, but much smaller than $H_I^2$, $0<C\ll 1$. This 
happens either in supergravity models  that possess a Heisenberg symmetry 
in which supersymmetry breaking by the inflationary vacuum energy does not 
lift flat directions at tree level \cite{heis}, or in models of hybrid
inflation based on orbifold constructions, in which a modulus field $T$ is 
responsible for the large value of the potential during inflation, and a 
second field $\phi$ with appropriate modular 
weight is responsible for the roll-over \cite{casasgelmini}. The correlation 
length for de Sitter fluctuations in this case is 
$\ell_c=H_I\:{\rm exp}(3H_I^2/2 m^2)$. This is large compared to the horizon 
size. In fact, the present length corresponding to $\ell_c$ should be larger 
than the horizon size today. Using $H_I=10^{13}$ GeV, this gives 
$(m/H_I)^2\lsim  40$, which is easily satisfied. The flat direction is 
truly flat and the AD baryogenesis may be implemented as originally 
envisaged \cite{affleckdine}.  
After inflation, these truly flat directions generate a large baryon 
asymmetry,  typically $B={\cal O}(1)$. Mechanisms for suppressing this 
asymmetry to the observed level have been considered in \cite{heis}. These 
include dilution from inflaton or moduli decay, and higher dimensional 
operators in both GUT models and the MSSM. The observed BAU can easily be 
generated when one or more of these effects is present.

\section{Conclusions}

The origin of the baryon to entropy ratio of the universe is one of the
fundamental initial condition challenges of the standard model of
cosmology. Over the past several decades, many mechanisms utilizing quantum
field theories in the background of the expanding universe have been proposed
as explanations for the observed value of the asymmetry. In this talk I have tried 
to give a brief overview of GUT scenarios, electroweak baryogenesis, leptogenesis and
the Affleck-Dine mechanism. Each of these mechanisms has both attractive
and problematic aspects.

While GUT baryogenesis is attractive, since the Sakharov criteria are so naturally satisfied,
it is not likely that the physics involved will be directly testable in the foreseeable future. 
While we may gain indirect evidence of grand unification with a particular gauge group, direct
confirmation in colliders seems unrealistic. A second problem with GUT
scenarios is the issue of erasure of the asymmetry - unless a GUT mechanism generates an
excess $B-L$, any baryonic asymetry produced will be equilibrated to zero
by anomalous electroweak interactions. While this does not invalidate GUT 
scenarios, it is a constraint. For example, $SU(5)$ will not be suitable
for baryogenesis for this reason, while $SO(10)$ may be.  

In recent years, perhaps the most widely studied scenario for generating 
the baryon number of the universe has been electroweak baryogenesis. The physics
involved is all testable in principle at realistic colliders and, furthermore,
the small extensions of the model involved to make baryogenesis successful
can be found in supersymmetry, which is an independently attractive
idea. It is worth stressing, however, that electroweak baryogenesis does not rely on
supersymmetry.

The testability of electroweak scenarios also leads to tight 
constraints. At present there exists only a small window of parameter space in
extensions of the electroweak theory in which baryogenesis is viable. This is because electroweak baryogenesis to be effective requires a strong enough first order phase transition. This translates into a severe upper bound on the lightest Higgs boson mass; $m_h \lsim 120$ GeV, in the case in which the mechanism is implemented in the MSSM. 
In addition, the stop mass must be smaller than, or of order of, the top quark mass.

If the Higgs and stop are found at the LHC, crucial tests will
come from tests of the requisite $CP$-violation through $B$ physics experiments and 
precision measurements of the Higgs and sfermion sectors at a future linear collider .

Affleck-Dine baryogenesis is a particularly attractive scenario, and much
progress has been made in understanding how this mechanism works. As was the
case for electroweak baryogenesis, this scenario has found its most promising 
implementations in supersymmetric models, in which the necessary flat 
directions are abundant. Particularly attractive is the fact that these
moduli, carrying the correct quantum numbers, are present even in the MSSM.

The challenges faced by Affleck-Dine models are combinations of those faced
by the GUT and electroweak ideas. In particular, it is necessary that $B-L$
be violated along the relevant directions (except perhaps in the Q-ball
implementations~\cite{k,ks,qb1,qb2,bbbd}) and that there exist new physics at scales above the
electroweak. 

No discussion of baryogenesis is complete these days without mentioning leptogenesis. This 
idea has found new support in recent years because of the discovery of neutrino masses. If these
are due to a heavy right-handed neutrino, then leptogenesis becomes a compelling mechanism.

Whatever the answer, the baryon asymmetry of the universe is clear cosmological 
evidence for new and interesting physics beyond the standard model of particle
physics. With the LHC imminent, a host of impressive B-physics experiments and a 
linear collider in the future, the prospects for making progress on this fundamental problem 
seem good.

\begin{acknowledgments}
I would like to thank the directors of SSI2004, JoAnne Hewett, John Jaros, Tune Kamae and Charles Prescott, for the invitation to lecture, for their hospitality and for an extremely stimulating time at SLAC. I would also like to thank Antonio Riotto for permission to draw heavily on our joint review article~\cite{Riotto:1999yt} in fleshing out the details of the actual lecture I gave at SLAC.
This work is supported in part by the NSF under grants PHY-0094122 and PHY-0354990, and by a Cottrell 
Scholar Award from Research Corporation.
\end{acknowledgments}

\end{document}